# An exploratory study of skill requirements for social media positions: A content analysis of job advertisements

Amit Verma*, Phillip Frank, Kamal Lamsal

Missouri Western State University

*Corresponding Author Email: averma@missouriwestern.edu Address: 4525 Downs Drive,

Saint Joseph, MO, 64507, Tel: +1 816 271 4357, Fax: +1 816 271 4338

**Abstract**

There has been considerable debate about the comparative advantages of marketing education emphasizing theoretical knowledge and applied skills. The current study investigated the skills necessary for entry-level marketing positions, specifically that of Social Media Manager (SMMgr) and Social Media Marketer (SMMkt). Data was collected from Indeed.com using a web crawler to extract job postings for SMMgr and SMMkt.  A total of 766 and 654 entry-level jobs for SMMgr and SMMkt, respectively, across the entire United States, was collected. Independent raters separately analyzed the data for keywords and categories. Findings suggest that the most desired skills are occupational digital marketing skills. Other relevant skill categories included communication, employee attributes, problem-solving, and information technology skills. This study extends the current literature by highlighting the desired skills prevalent across the social media industry. The findings also have relevance in designing the marketing education curriculum, specifically in isolating core skills that could be integrated into the marketing courses.

## 1. INTRODUCTION

There has long been debate about the comparative advantages of marketing education emphasizing theoretical knowledge or applied marketing skills. (Walker et al. (2009); Finch, Nadeau, & O'Reilly (2012)). Those arguing for theoretical knowledge stipulate that technical skills tend to be domain-specific and that while these skills may change across domain and time, that the conceptual knowledge remains more stable and may serve as the foundational structure for technical skills. Proponents of technical skills, on the other hand, argue that students entering into the workforce need more applied skills that may provide benefits for students from day one (Barr & McNeilly (2002); Schlee & Harich (2010)). In their 2002 article, Barr and McNeilly interviewed recruiters and found that internships and part-time positions offered a better indication of a student's employability. They also emphasized the need for more applied learning opportunities in classrooms to address technical skill gaps.

Social media skills have become increasingly important for success in the digital economy. In fact, social media positions have been recognized as one of the fastest-growing job categories (Hallett (2016) and Moore (2017)). Thus, the current study looks to investigate the key skills necessary for specific entry-level marketing positions in the social media landscape. We focus on entry-level positions due to the thesis that most undergraduate students are more likely to move into these positions over middle-level and upper-level ones. While there are cases where a student may achieve a higher-level position directly after finishing undergraduate education, these scenarios appear as outliers and do not serve as a fair representation of the majority of marketing graduates. Thus, our study conducted a content analysis of social media job postings for entry-level marketing positions posted between June and December of 2019 on Indeed.com. We specifically looked at postings for Social Media Manager (SMMgr) and Social Media

Marketer (SMMkt). Our study seeks to extend the work of previous scholars and to provide a fresh perspective on the prominent skills desired for new hires within the social media niche in the digital marketing field.

## 2. HYPOTHESIS DEVELOPMENT

In their 2010 article, Regina Schlee and Katrin Harich investigated the skills and conceptual knowledge required for marketing positions at all levels. The data was collected via a content analysis of 500 positions on Monster.com across five major cities in the US (Atlanta, Chicago, Los Angeles, New York City, and Seattle). The authors' findings support an emphasis on technical skill development as part of marketing education for more entry-level positions and conceptual skills such as developing a marketing plan, managing marketing functions or supply chain design being more prevalent for "Upper-level" (5+ years) management positions. The authors conclude that marketing education programs should address more skill-specific areas of focus and conceptual knowledge.

In their updated article written in 2017, Regina Schlee and Gary Karns further articulate the importance of specific skills sets including analytical skills, project management, and statistical software skills such as Excel, SPSS, and Google Analytics showed a high frequency of denotation in entry-level positions and showed an initial correlation to the expected salary of job postings. Furthermore, more general skills such as oral communication, presentation skills, and time management also showed high importance in their search. The authors in both articles conclude that experience-based learning opportunities such as applied learning experiences, enable students to gain an added advantage in preparedness in their marketing career pursuits. This finding was echoed in the 2018, Douglas and Randall Ewing article that posits that marketing education curriculums should be designed to reflect more of a "working professional" identity where classrooms become more experiential in construction and where learning is more applied.

Even with a growing assemblage of authors arguing for the limitation or all out removal of theory from marketing education, there is no denying the importance of theoretical knowledge in the development of more marketable graduates. Wellman (2010) argued that the inclusion of theory in marketing education is directly related to the establishment of higher-order cognitive skills in student development. He further articulates that these higher-order cognitive skills include imagination and creativity, adaptability, and continuous desire to learn, all of which are highly prized in marketing careers. It were these broader skills that could be transferred across various marketing positions and industries that served to solidify a graduate's career success more so than position-specific skills.

It is in this same vein that Victoria and William Crittenden wrote their 2015 article that investigated the role played by technology in contributing to the ever-changing marketing discipline. Specifically, the authors argued that given how quickly technology changes as well as the increased role that digital media has on marketing strategy, that educators must develop curricula that emphasizes the general theory of marketing thought as applicable to the digital medium. The focus here is again on transferrable knowledge that graduates may apply across the multitude of digital media platforms both currently available as well as those yet to be established.

Given the high job demands in the social media landscape, we believe that an in-depth exploration of the critical skill requirements of the social media positions is warranted. Thus, we could compare the importance of theoretical and applied skills from the employers' perspective. To the best of our knowledge, the skill requirements for these specific social media jobs have not been explored in the literature. Hence, the current study looks to investigate the skills desired by the firms in the United States for entry-level Social Media Manager and Social Media Marketer positions.

## 3. LITERATURE REVIEW

Social media job roles have risen into prominence in the last decade, with SMMgr and SMMkt positions peaking in demand. Hallett (2016) and Moore (2017) touted SMMgr as a job position that was non-existent a decade ago. However, with the digital revolution, the interest in digital marketing has fueled the growth in this vertical. The authors argued that online platforms like Facebook, Instagram, and Twitter provide an indispensable marketing opportunity for product branding and growth. Moretti & Tuan (2015) define SMMgr as a middle-level managerial position that has swiftly evolved as a new role in the management of social media dealing with the organization's reputation. The authors argue that a team comprising of Social Analyst, Content Manager, Social Media Strategist, and Community Manager is essential for the execution of any social media strategy. Bon (2014) traced the evolution of marketing manager into social media manager. The author also identified the internal and external factors that influence a firm's social media adoption and studied the practical characteristics of SMMgr through empirical analysis of surveys sent to Hungarian and Italian companies. On the other hand, our study focuses on the social media employer's perspective in the US via job advertisements.

The impact of social media on various business functions has been well known. Pourkhani, Abdipour, Baher, & Moslehpur (2019) studied the applications of social media in business through a scientometric analysis of published articles between 2005 and 2019. Their analysis produced theme clusters that can be used to identify specialized niches in this domain. Outside business, a field that has seen considerable impact of social media is mass media or journalism, which has significantly evolved with the advent of social media tools. Regarding skill development, the authors in Van Dijk & Van Deursen (2010) present the similarities and differences between traditional mass media and digital media skills. They argue that the latter puts more emphasis on information technology and strategic decision making. They classified Internet skills into operational, formal, information, and strategic categories. Moreover, they also provide a framework to measure digital media skills. The authors in Wenger & Owens (2012) conducted a content analysis of all employment opportunities posted by the newspaper and journalism companies for specific time periods in 2008 and 2009. They coded the required attributes and skill categories for journalism jobs. The authors acknowledge an increased impact of social media skills and content management on these jobs. The aforementioned digital marketing skill gap has also been recognized in communication industries. Royle & Laing (2014) found a lack of specific technical skills like video editing, animation, website development, code writing, etc. following an in-depth interview of 20 industry professionals.

The importance of social media skills in education has been well documented in the literature.  For instance, Lu, Newman, & Miller (2014) studied students' perceptions regarding social media as a learning tool. They provide various recommendations for educators on the integration of social media into the academic curriculum. This scope of curriculum development is also explored in Novakovich, Miah, & Shaw (2017). The authors designed a social media component for a professional writing course after identifying gaps between students' social media usage and professional practice. The authors concluded that the students require significant guidance to tailor their social networking presence towards professional jobs. This skill gap is also addressed in Benson, Morgan, & Filppaios (2014). The authors report that UK business graduates are not well equipped with the social networking skills necessary for career growth. From the results of a questionnaire-based study, they conclude that social networking awareness is missing in the business curricula.

The related skill research in the marketing field could be classified with respect to the point of view of employees and employers. The former's emphasis is on the perceived skills required to succeed in the workplace. For example, Böttcher (2019) explored the skills required for marketing positions in Netherlands. The author also identified the factors influencing each skill. The results of the study comprising of 150 survey responses showed that the positions required average levels of information management, critical thinking, creativity, and collaboration skills. Raghuraman (2017) examined the skills necessary for entry-level digital marketing positions. The results from twelve in-depth interviews suggest that in the digital marketing domain, verbal and written communication skills are more valuable compared to technical or data analytical skills. The authors in Sutherland & Ho (2017) linked the graduate employability with social media skills. This study explored the perception of undergraduates in medicine, law, science, and arts towards social media pedagogy.

From the employer's perspective, various studies have been conducted to analyze the required skills. For example, Barker (2014) developed a relative importance rating for employability skills through qualitative research with employers. The communication skill was seen as the most important skill followed by problem-solving and personality traits like enthusiasm, self-confidence, self-motivation, maturity, and willingness to learn.

Due to the advent of big data and the proliferation of online platforms for job advertisements, it is easier to collect and analyze multiple data sources related to employment using content analysis. The employer's perspective in the general marketing domain has been studied using data from job sites like Indeed, Monster, LinkedIn, etc. For instance, Schlee & Harich (2010) analyzed 500 marketing jobs from metropolitan areas in the US like Atlanta, Chicago, Los Angles, New York City, and Seattle. The source was Monster.com, and they differentiated entry-level, lower-level, middle-level, and upper-level positions. The findings suggest that for all job types, technical skills like MS Office, statistical software, database analysis, data mining, internet marketing tools, etc. were more important than conceptual marketing knowledge. Rosenstreich, Priday, & Bedggood (2014) investigated 729 job advertisements for recent marketing graduates from Australian and US job sites. The study found that the skills corresponding to data analysis and communication were most prominent. The personality attributes of confidence, ambition, and determination were also desired by employers. A similar analysis was carried out by authors in McArthur et al. (2017) using 359 job advertisements posted on the job website seek.com.au in Australia. They established that employee attributes like motivation and time management were most demanded in addition to communication and digital marketing experience. More recently, the authors in Schlee & Karns (2017) examined the skill requirements and personality attributes for 210 entry-level job listings in the US from Indeed.com and LinkedIn.com. The importance of analytical and technological skills was highlighted in addition to core marketing principles. From the European job market perspective, the authors in Gregorio, Maggioni, Mauri, & Mazzucchelli (2019) also employed content analysis to study 359 job advertisements in Italy, France, Germany, Spain, and United Kingdom. They developed a framework defining five skill categories and 29 skills for marketing professions. Their study revealed that basic soft skills are the most valuable for marketing professionals. They also found that analytical skills are highly regarded.

Instead of focusing on all existing entry-level marketing jobs, our current research seeks to contribute to the existing literature by identifying the skill requirements for two popular job titles in the social media landscape: SMMgr and SMMkt. We conduct a pairwise comparison between SMMgr and SMMkt. Thus, the findings of this research aim to shed light on relatively unexplored categories of social media jobs. Due to the advent of social media, these jobs have increased in popularity (Hallett (2016) and Moore (2017)). The present study also contributes to the existing literature by introducing new skill categories required for social media positions. The findings could be very beneficial for redesigning the existing digital marketing curricula.

## 4.   RESEARCH METHODOLOGY

### 4.1. Skill Classification Framework

In this section, we present the different skill categories used for job postings. Our objective is to classify whether each job posting belongs to one or more of these skill categories based on the keywords. We used a list of skill categories developed by McArthur, Kubacki, Pang & Alcaraz (2017) as a starting point and expanded it using a more recent framework developed by Gregorio et al (2019). Also, we added more skill categories to reflect the needs in the social media landscape. For achieving this, we took a sample of 50 jobs of SMMgr and SMMkt and diligently built the keywords list necessary for social media utilizing inputs from McArthur et al. (2017) and Gregorio et al. (2019). Each of these skill categories and subcategories was carefully analyzed and updated based on the authors' inputs. We use a set of three independent raters to measure the reliability score of our categorization structure. The alpha coefficient of the intercoder reliability measure based on Kim & Lee (2016) was 0.93. We present the final list of categories and subcategories along with sample keywords in Table 1.

*Table 1*: Classification framework – skill categories and associated skills

| Skill Category | Skills | Keywords |
|---|---|---|
| Communication | Written | Copywriting, Editing, Blogging, Content Creation, Story-ideation |
| | Verbal | Verbal, Oral, Cold calling |
| | Presentation | Present, Presentation, Report |
| | Generic | Responsible, Determined, Competitive, Witty, Go-getter, Success-oriented |
| Employee Attributes | Motivation | Motivated, Ambition, Willingness to learn, Delivering result, Continuous learning |
| | Time Management | Time management, Timely manner, Prioritize time, deadline driven |
| | Detail oriented | Attention to detail, Eye for detail, Accuracy, Precision |
| | Attitude | Can do, Self-learner, Self-directed, Positive Attitude |
| | Independence | Independence, Without supervision, Autonomous |
| | Adaptability | Adaptable, Flexible, Multitasking |
| | Confidence | Confident, Decisive |
| | Other | Funny, Smiling, High energy, Reliable, Proactive |
| Occupational Attributes | Digital Marketing | Online reputation, Digital engagement, Reputation management, Performance monitoring, Community outreach, Content marketing |
| | Project Management | Project management, PERT, CPM, PERT/CPM, change management, project budget, project documentation, PMP, Microsoft Project, Gannt Chart |
| | Generic | Print marketing, Advertisement, Target market, Brand management, Viral marketing, Product awareness, Segmentation, Marketing channel, Consumer behavior, Marketing strategy, Promote content |
| | Campaign Management | Develop campaign, Manage outreach, Public relations, Generate buzz |
| | Sales | Sales, Create revenue, Lead generation, Drive sales |
| | Customer Service | Customer service, Service level agreement, customer experience |
| | B2B | Business To Business |
| | Domain | E-commerce, Finance, healthcare, marketing, supply chain, accounting, computer science, functional, domain |
| Information Technology | MS Office | Excel, Word, Powerpoint, Outlook, Office |
| | Adobe | Photoshop, Illustrator, Indesign, Lightroom, Dreamweaver |
| | Other | Coding, Programming, Videography, Photography |
| | Web Design | HTML, CSS, XML, Hashtag, Website design, Home page |
| | Analytics | Google Analytics, SPSS, SAS, Big data, R |
| | Social Media Tools | Sprinklr, Percolate, SEO, SEM, Sprout Social, Spredfast, Trello, Facebook business manager, Hootsuite, Shopify |

| Interpersonal | Interpersonal | Team management, Collaboration, Cooperation, Networking, Client relationship |
|---|---|---|
| Problem Solving | Problem Solving | Problem solving, Troubleshoot, Conflict resolution, Solve issue, Critical thinker |
| | Creativity | Creative, Out of box, Storyteller |
| | Process Design | Design Process, Improve process, Continuous improvement, Operations management |
| Administrative | Administrative | Issue management, Posting schedule, Product launch, social calendar |
| Research | Research | Data gathering, Data collection, Data reporting, Monitor trend, Monitor performance |
| Numeracy | Numeracy | Financial Management, Bookkeeping, Accountancy |
| Foreign Language | Foreign Language | Spanish, French, German, Italian, Chinese |

### 4.2 Research Method

For this project, we used data from Indeed.com, the most popular job search site across the US. Web scraping was the preferred method of data collection. The involved time phase was June – December 2019. We performed a search on Indeed.com based on the titles of "Social Media Manager" and "Social Media Marketing." After downloading the job postings using our web scraper, the job description was broken down into keywords using content analysis. Content analysis is a popular technique to study words in a document and identify patterns (Neuendorf (2016)). The text needs to be broken down into categories to summarize data. Recall that each of these keywords belongs to a specific category and subcategory according to our skill classification framework described in the previous section. This enables us to quantify the required skills for each job posting. Thus, we could build aggregate measures for relative importance of each skill category with respect to these two job titles. We highlight the similarities and differences between these two job titles in the next section. Our technique is summarized in Figure 1.

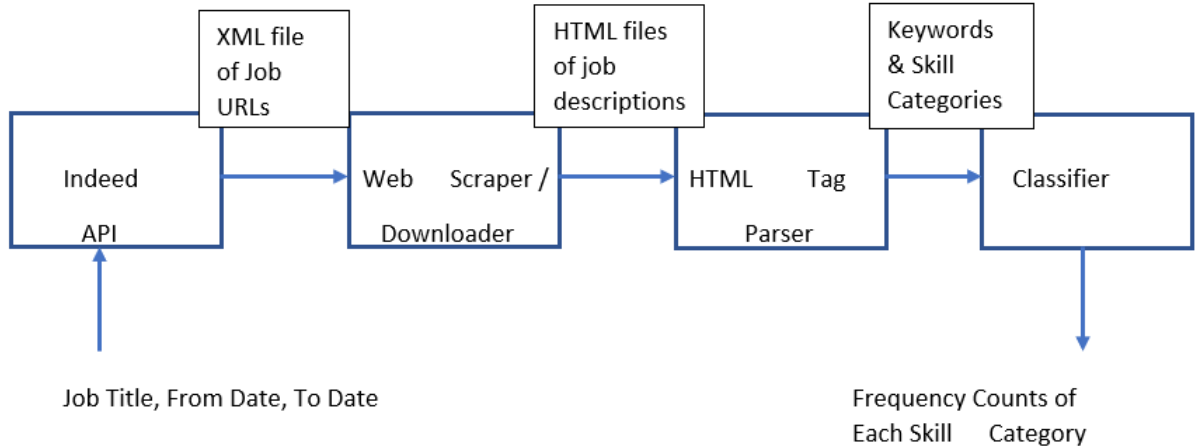

*Figure 1: Summary of our keyword extraction technique*

Next, we describe each component of our method. First, we start with the web scraper. We used Python to develop the web scraping tool. It is heavily reliant on the existing Indeed API. The API takes job titles, location, and time period as input arguments. We then used the query phrases of "Social Media Manager" and "Social Media Marketing" during the specific time period in the United States. The output is an XML file containing the job title, job URL, location, company, posting date, and a job summary. The job URL is very important for our analysis since the job summary field of the API lacks necessary information like programming tools, statistical packages, and web design experience. Hence, we downloaded the actual job postings using the URLs specified by the API.

The downloading operation was also performed programmatically, and the files were stored locally. The stored HTML file has different attributes stored under different tags. However, the benefit of using Indeed is that it uses a consistent design for these webpages. Hence, the job descriptions are always stored using a specific tag. This simplifies the development of the parser, which extracts the job description for each downloaded HTML file. The job description field is important since it contains information like position type, salary, academic experience, and required software tools. Further, the parser removes all unnecessary keywords like a, the, in, for, and special symbols like quotation marks and colon. This practice of removing stop words is standard in natural language processing used for content analysis (Neuendorf (2016)).

The job description field is the input to a statistical tool that counts the frequency occurrence of each category and subcategory for the two job titles. The associated keywords for each skill category are listed in Table 1. We performed a match between keywords belonging to these skill categories and the keywords present in the job description. If a match was found, we declared that the associated skill category is required for a specific job posting. This process is repeated for each job posting and each skill category. Aggregating all the results, we counted the number of jobs for which a specific skill category is required. This represents the frequency of occurrence of each skill category for each job title. Using this information, we could establish the relative importance of each skill category for each job title. We utilized the same for comparative analysis in Tables 4-5 of the Results section.

### 4.3 Results

We collected a total of 766 and 654 entry-level jobs for SMMgr and SMMkt, respectively, from Indeed.com in the time period of June-December 2019. The sample size is big enough to qualify as a representative sample for statistical analysis. The distribution of the job postings with respect to the US states for SMMgr is presented in Figure 2. Note that the location information is part of the Indeed API output for different job titles. The top five states are CA, NY, TX, FL, and IL. This is explained by the concentration of the population centers and the marketing industries. The job positions for SMMkt exhibit similar behavior.

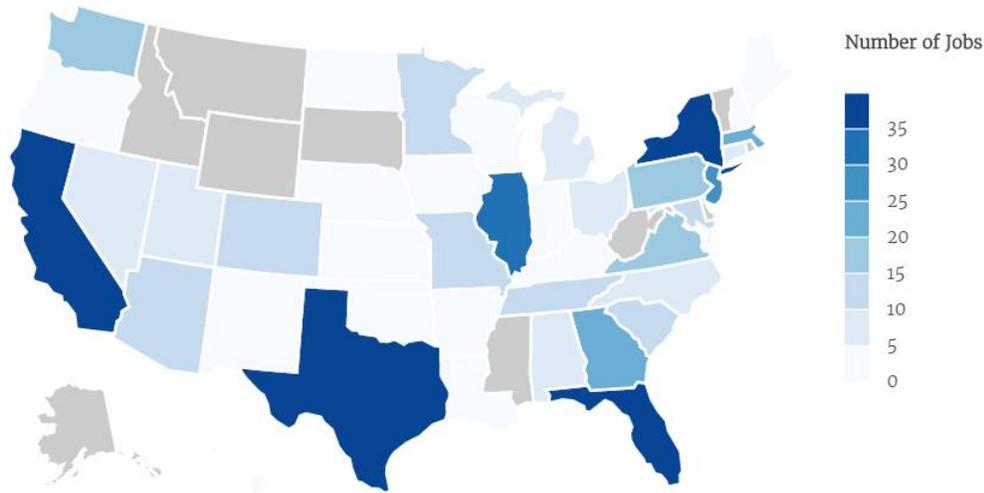

*Figure 2: Geographical Distribution of Social Media Manager Jobs*

For these entry-level positions, we parsed the information related to majors in addition to the specific keywords listed in Table 1. This was accomplished by regular expressions using the job descriptions. We were interested in scenarios where specific majors like Marketing, Communication, etc. were present in the vicinity of contextual information like Major, Minor, BS, BA, etc. If such a scenario existed, we declared that a specific major was required for a job posting. Next, we developed a frequency count for each major in a similar manner as job categories. The distribution across different majors for SMMgr is presented in Table 3.

*Table 3:* Distribution of Majors

| Major | Jobs % |
|-------|--------|
| Marketing | 87.08 |
| Business | 55.22 |
| Commerce | 6.14 |
| Property | 3.92 |
| Quantitative (Finance /Mathematics) | 3.13 |
| Engineering | 1.83 |
| Economics | 1.17 |

The finding that the top required majors are Marketing / Business is obvious. The next majors in the list are Communications, Commerce, and Property. These degree programs are sometimes niche in the marketing domain and have a significant correlation with traditional marketing offerings. The other related degree programs are Finance and Economics. Again, SMMkt job positions exhibit a similar trend.

Next, we present an in-depth analysis of SMMgr and SMMkt job positions with respect to associated skill categories. This analysis can be conducted in two ways. First, we could focus on each job title like Boyle & Strong (2006) and Kim & Lee (2016). Second, we could compare the two job titles based on the associated skill categories similar to Aasheim et al. (2015), Verma et al. (2019) and Verma et al. (2021). We used the latter framework for our analysis. In this way, we could compare the two job titles analytically.

We present the relative importance of each skill category for SMMgr and SMMkt in Tables 4 and 5 respectively. The relative frequency with respect to the total number of jobs is presented in the column Percentage Count (%). Note that a number of 20% means that the specific skill subcategory is required in 20 out of 100 job postings. Also, the percentage counts for each skill category do not add up to 100%. These skill categories are sorted with respect to the total number of jobs that require the expertise in a particular skill category. Further, we report the top five skills corresponding to each skill category. For instance, the skill Digital Marketing is present in 99% of job advertisements corresponding to SMMgr. Thus, at least one or more of the keywords listed in Table 1 for the skill of Digital Marketing are present in 99% of the job advertisements. This is expected behavior since both these job positions utilize a majority of digital marketing skills.

*Table 4*: Social Media Manager

| Skill Category | Skill | Percentage Count (%) |
|---|---|---|
| Occupational Attributes | | |
| | Digital Marketing | 99.05 |
| | Generic | 78.33 |
| | Communication | 53.52 |
| | Project Management | 24.41 |
| | Sales | 24.15 |
| Communication | | |
| | Written | 78.07 |
| | General | 77.02 |
| | Verbal | 65.80 |
| | Presentation | 23.63 |
| Employee Attributes | | |
| | Time management | 74.15 |
| | Motivation | 50.13 |
| | Attention to Detail | 42.82 |
| | Adaptability | 40.99 |
| | Work Ethics | 33.94 |
| Problem Solving | | |
| | Creativity | 66.84 |
| | Problem Solving | 43.47 |
| | Other | 16.06 |
| Information Technology | | |
| | Other | 54.40 |
| | Analytics | 45.30 |
| | Social Media Tools | 41.64 |
| | Web Design | 24.28 |
| | Adobe | 22.98 |
| | MS Office | 17.10 |

*Table 5*: Social Media Marketer

| Skill Category | Skill | Percentage Count (%) |
|---|---|---|
| Occupational Attributes | | |

| | Digital Marketing | 98.78 |
|---|---|---|
| | General | 75.99 |
| | Communication | 40.52 |
| | Sales | 33.94 |
| | Customer Service | 16.21 |
| Communication | | |
| | Written | 76.15 |
| | General | 69.27 |
| | Verbal | 62.84 |
| | Presentation | 14.22 |
| Employee Attributes | | |
| | Time management | 66.51 |
| | Motivation | 54.13 |
| | Attention to Detail | 42.66 |
| | Independence | 41.74 |
| | Adaptability | 39.91 |
| Problem Solving | | |
| | Creativity | 61.93 |
| | Problem Solving | 26.45 |
| | Process Design | 13.76 |
| Information Technology | | |
| | Other | 51.00 |
| | Web Design | 37.61 |
| | Analytics | 32.11 |
| | Social Media Tools | 30.28 |
| | Adobe | 29.36 |
| | MS Office | 25.38 |

The next step is to establish whether there is a statistically significant difference between the two job titles based on the skill subcategories. For this purpose, we compared the percentage count of each skill subcategory for SMMgr and SMMkt. In total, there are 40 skill subcategories. We ran a hypothesis t-test based on paired two sample for means. The resulting p-value of 0.015 with a significance level of 5% led us to reject the null hypothesis that the two population means are equal. Hence, the two paired sets of percentage counts for SMMgr and SMMkt are different, with a statistical significance level of 5%.

From Tables 4 and 5, we observe that occupational attributes like digital marketing are very crucial for SMMgr and SMMkt. These findings closely match the results of Schlee & Karns (2017) for generic marketing positions. However, the percentage emphasis on digital marketing skillset is much more. The most important skills related to social media are managing web traffic, content marketing, performance tracking, and measuring social media performance using metrics. The generic subcategory is the second most popular, including traditional marketing tasks related to the management of target market, lead generation, brand marketing, and advertising. The other important skill of digital communication involved strategies tied to marketing campaigns and public relations. This finding emphasizes that core marketing skills related to the target market, brand marketing and advertisement design are still crucial for the digital economy's social media jobs. Project management skills are more important for SMMgr. As the job title suggests, SMMgr requires more managerial skills. On the other hand, this skill category does not appear in the top-five list for SMMkt. The traditional roles of marketing, such as sales and customer service, are of lower significance for both SMMgr and SMMkt.

The second most important skill category of communication is typically considered as soft skills. Lately, they have become increasingly important for all job types. Compared to a traditional sales job where verbal skills are in demand, social media jobs require a lot of copy editing. The typical activities involve blog creations, posting, social media messaging, etc. Hence, the written skills hold more significance for these two job titles. The generic communication skills involving attributes like team leadership, determination, ambition, and determination are necessary for career growth in business verticals like this.

The next category of importance is the personality traits related to the employee. These include time management, motivation, independence, attention to detail, and adaptability. Time management skills are very crucial, especially for the millennial workforce. This has been corroborated by various studies in the human resource management literature (see Howe & Strauss (2007)). These findings are consistent with the results for entry-level marketing positions achieved by McArthur et al. (2017). The employees are also expected to be multi-tasker, well organized, and detail-oriented without the need of additional supervision. The companies are interested in candidates who can perform well on their own in fast-paced, high-stress, deadline-driven environments. Further, employees should be flexible and capable of handling multiple projects simultaneously.

Problem-solving skills have become increasingly important for the modern workforce. These include critical thinking, innovation, and creativity. The candidates should possess sound judgment abilities. They should be able to think clearly and rationally while solving a business problem. The creativity attribute deals with thinking out of the box, which is tied with visual storytelling. In the age of social media advertising, fresh content creation is crucial. The other skills in this category deal with process and continuous improvements and holistic process design. Thus, knowledge of supply chain and operations management constructs are somewhat important for social media marketing jobs. This is because marketing channels have started gaining traction in this global era with interconnected supply chains.

The other important category concerns information technology skills, which are usually considered as hard skills. The biggest subcategory are the generic tools like coding, programming, videography, and photography. The two job positions involve curating photos and videos for the social media feed along with basic knowledge of editing tools.  It is also worth noting that the web design tools are more in demand for SMMkt when compared to SMMgr. As discussed earlier, SMMgr positions are managerial, while SMMkt job roles are somewhat technically oriented. The web design skillset requires familiarity with HTML, CSS, XML, CANVA, etc. More specifically, the job positions entail activities like designing websites, home page, landing page, etc. In terms of content creation, it involves curating new graphics, layouts, and tags. The analytical software include Google Analytics, R, SPSS, Excel, and Hootsuite.

We present the distribution of social media tools with respect to their relative frequency of occurrence in the SMMgr job advertisements in Table 6. We could clearly observe from the results that the top five software tools in this category are Search Engine Optimization / Marketing (SEO / SEM), Hootsuite, Sprout Social, Sprinklr, Buffer, and Facebook Business Manager. SMMkt job positions follow a similar trend. This is an important insight in terms of curriculum design. The digital marketing course curriculum should be updated so that more emphasis is placed on these software packages. More specifically, SEO / SEM should be incorporated in the marketing curricula by using these software packages. Lastly, Microsoft Office and Adobe tools like Excel, Illustrator, Dreamweaver, Lightroom, etc. have continued their prevalence in the hard skills need to succeed in the workplace.

*Table 6*: *Frequency Distribution of Social Media Tools*

| Social Media Tools | Percentage Count (%) |
| --- | --- |
| SEO / SEM | 23.75 |
| Hootsuite | 11.09 |
| Sprout Social | 6.78 |
| Sprinklr | 6.39 |
| Buffer | 2.87 |
| Facebook Business Manager | 2.74 |

## 5. CONCLUSIONS AND FUTURE RESEARCH

In this paper, we investigated the job advertisements for two popular job titles in the social media landscape: Social Media Manager and Marketer. We conducted a nationwide study across the US using the scraped jobs from Indeed.com. Further, we performed a pairwise comparison between SMMgr and SMMkt, highlighting the similarities and differences. We found that the most desired skills are occupational digital marketing skills. Other relevant skill categories like communication, employee attributes, problem-solving, and Information Technology skills are explored in detail.

The SMMkt appears to be somewhat more technical compared to SMMgr with an emphasis on web design skills. Moreover, project management skills were more valued for SMMgr, highlighting that these jobs are upper management jobs involving strategic decision making. The current study presented a sorted ranking of the skills required for social media positions. Further, we discussed the relative significance of the skills with respect to the frequency of occurrence in the total job advertisements.

The findings of this study could be utilized in two different ways. First, they could be used by the human resources team in the industry to provide concise and well-defined job descriptions for the two job titles. This would save a lot of money and resources for people who are in charge of formulating the job advertisements. Second, the business curriculum redesign efforts can be directed using the findings of this research, especially given the prevalence of the digital economy and the advent of social media. More specifically, digital marketing programs could be redesigned to reflect the current trends in the workplace with respect to the in-demand social media tools and underlying principles.